\documentclass[11pt,a4paper]{article}
\usepackage[left=25mm,right=25mm,top=25mm,bottom=30mm]{geometry}
\usepackage[T1]{fontenc}
\usepackage[english]{babel}
\usepackage{lmodern,amsmath,amssymb,amsthm,mathtools}
\usepackage{microtype,setspace,booktabs,tabularx}
\usepackage[shortlabels]{enumitem}
\usepackage{cite,lineno}
\usepackage[hidelinks]{hyperref}
\usepackage{needspace}
\usepackage{tikz}
\usetikzlibrary{arrows.meta,calc}
\usepackage[section]{placeins}
\theoremstyle{plain}
\newtheorem{theorem}{Theorem}
\newtheorem{lemma}[theorem]{Lemma}
\newtheorem{corollary}[theorem]{Corollary}
\newtheorem{proposition}[theorem]{Proposition}
\theoremstyle{definition}
\newtheorem{definition}[theorem]{Definition}
\theoremstyle{remark}
\newtheorem{remark}[theorem]{Remark}

\newcommand{\NN}{\mathrm{NN}}
\newcommand{\HNN}{\mathrm{HNN}}
\newcommand{\MP}{\mathrm{mpPTF}}
\newcommand{\RES}{\mathrm{RES}}
\newcommand{\dist}{\operatorname{dist}}
\newcommand{\Q}{\mathbb Q}
\newcommand{\R}{\mathbb R}
\newcommand{\Z}{\mathbb Z}
\newcommand{\N}{\mathbb N}
\newcommand{\cube}{\{0,1\}^n}
\newcommand{\ip}[2]{\langle #1,#2\rangle}
\newcommand{\norm}[1]{\lVert #1\rVert_2}
\newcommand{\poly}{\operatorname{poly}}

\title{Local Modifications of Nearest-Neighbor\protect\\
and Min-Plus Representations}
\author{%
Liliia Redina\\[3pt]
\small HSE University, Moscow, Russia\\
\small\href{mailto:liliiaredina@gmail.com}{\texttt{liliiaredina@gmail.com}}}
\date{}
\hypersetup{pdftitle={Local Modifications of Nearest-Neighbor and Min-Plus Representations},
pdfauthor={Liliia Redina}}

\begin{document}
\maketitle

\begin{abstract}
We study how changing a Boolean-valued function at $d$ inputs affects
the size and precision of its exact representations. Starting from a
representation of size $m$, we prove four bounds, two for nearest-neighbor
(NN) representations and two for integer min-plus representations.
\begin{enumerate}[label=(\roman*),leftmargin=*,itemsep=4pt,topsep=4pt,parsep=0pt]
\item For NN representations on a finite subset of $\R^n$ with $n\ge2$,
at most $m+2d$ anchors suffice. The coefficient $2$ is optimal:
on a three-point domain, changing one value can increase the minimum
number of anchors from one to three.

\item For NN representations on the Boolean cube with rational anchors,
let $R$ be the maximum original coordinate bit length, and let $\Delta$
be the maximum bit length of the least common denominator of any one
original anchor's coordinates. We construct a representation with at most
$m+3d$ anchors on a common dyadic grid and coordinate bit length
$O(R+\Delta+\log(n+1))$.

\item For integer min-plus representations on the Boolean cube, we give
a construction that keeps the original $m$ forms and their labels and
adds at most $d$ new forms. Its maximum weight is at most $W+2n$,
where $W$ is the largest absolute original coefficient, including
constant terms.

\item For integer min-plus representations on a finite rational domain
$X\subset\Q^n$, at most $m+2d$ forms suffice. Their maximum weight is
polynomially bounded in $m,d,W,n,|X|$, the coordinate bound, and the
common denominator of $X$. The coefficient $2$ is again optimal:
on a three-point domain, changing one value can increase the minimum
number of forms from one to three.
\end{enumerate}
These bounds imply stability under polynomially many changes on the
cube for polynomial-size representations, including those with
polynomially bounded NN coordinate bit length or min-plus weight.
\end{abstract}

\noindent\textbf{Keywords:} Boolean functions, nearest-neighbor complexity,
min-plus threshold functions, local modifications, rational resolution.

\section{Introduction}\label{sec:intro}

An exact representation of a Boolean function may be known even when
the function of interest differs from it on a small exceptional set.
This raises a basic question: how much additional descriptive
complexity is needed to correct those values? For representations by
nearest neighbors, the answer is not immediate. Adding an anchor can
change the classification of several inputs, and moving an existing
anchor can alter a substantial part of its Voronoi region.

We consider two models. In a nearest-neighbor (NN) representation,
finitely many points in Euclidean space carry binary labels, and an
input receives the label of its nearest anchor. In a min-plus
representation, the input is classified by comparing the minima of
two finite families of affine forms. Our aim is to control the cost of
modifying an existing representation: the number of anchors or forms,
and the number of bits needed to specify their coordinates or coefficients.
The input domain remains fixed throughout each modification.

\paragraph{Related work.}
Hajnal, Liu, and Tur\'an~\cite{Hajnal2022} systematically studied
nearest-neighbor representations of Boolean functions, distinguishing
real anchors from anchors restricted to the Boolean cube.
They determined the NN complexity of parity and proved upper and lower
bounds for several explicit functions. Their results show that
restricting anchors to Boolean points can substantially increase
the required number of anchors.

Kilic, Sima, and Bruck~\cite{KilicSimaBruck2023} studied both the number of
anchors and their resolution. For symmetric Boolean functions, they
constructed representations with one anchor per interval of constancy
and logarithmic resolution, and proved optimality of the number of
anchors when all intervals of constancy have the same length.
Their subsequent work examines
size--resolution tradeoffs for threshold functions~\cite{KilicSimaBruckNeurons2024v2}
and gives explicit NN constructions for several classes of neural
circuits and decision lists~\cite{KilicSimaBruckNN2024}. These studies
motivate quantitative results that control the numerical cost of
modifying a representation already available for a nearby function.

Min-plus polynomial threshold functions have been studied in relation
to threshold circuits~\cite{Hansen2015}. DiCicco, Podolskii, and
Reichman~\cite{DiCicco2024} characterized the closures of
polynomial-size NN and Boolean-anchor NN representations under
variable identification and substitution of constants. These closures
coincide with polynomial-size min-plus representations with
unrestricted and polynomially bounded integer weights, respectively.
Our constructions operate on the original domain and quantify the
cost of changing a specified set of values.

For Boolean circuits, Golovnev et al.~\cite{Golovnev2019} use the fact
that changing one truth-table entry changes the minimum circuit size
by at most $O(n)$. The resulting $O(nd)$ bound for $d$ changed entries
over a fixed finite complete gate basis is stated explicitly
in~\cite[Theorem~1]{Krinkin2026}.

\paragraph{Our results.}
Let $\dist_X(f,g)=|\{x\in X:f(x)\ne g(x)\}|=d$.
For nearest-neighbor representations we prove
\[
|\NN_X(f)-\NN_X(g)|\le2d
\]
for every nonempty finite $X\subset\R^n$ with $n\ge2$
(Theorem~\ref{thm:nn-local}). For each changed input, we relabel its
nearest anchor and add two anchors with the original label.
A three-point example shows that the constant
cannot be decreased over general finite domains.

Given an NN representation of $f$ on the Boolean cube by $m$ rational
anchors with coordinate matrix $A$, we construct a representation of $g$ with at most
$m+3d$ anchors and resolution
\[
O\bigl(R+\Delta(A)+\log(n+1)\bigr)
\]
(Theorem~\ref{thm:nn-resolution}). Here $R=\RES(A)$ is the maximum
coordinate bit length and $\Delta(A)$ is the maximum bit length of
the least common denominator within a single anchor; no denominator
common to different anchors is required. The proof separates all changed inputs
before correcting their labels, so precision does not accumulate
over the $d$ modifications. The output can use one common dyadic
denominator.

For integer min-plus representations on the Boolean cube we turn a
representation $\mathcal F$ of $f$ into a representation $\mathcal G$ of
$g$ satisfying
\[
|\mathcal G|\le |\mathcal F|+d,
\qquad
W(\mathcal G)\le W(\mathcal F)+2n
\]
(Theorem~\ref{thm:mp-cube}), where $W$ is the largest absolute
coefficient, including the constant term. We use the affine form
$2d_H(x,y)-1$, which is negative at the selected vertex $y$ and
positive at every other vertex.

On an arbitrary finite rational domain, Theorem~\ref{thm:mp-weight}
modifies an integer min-plus representation of size $m$ using at most
$m+2d$ forms and gives a polynomial bound on their weights. The
construction orders the inputs of each cell by a scalar projection and
assigns one form to each run of the target labels. It follows that the
minimum size changes by at most $2d$, with equality possible even on a
three-point domain (Corollary~\ref{cor:mp-local}).

Sections~\ref{sec:nn-local} and~\ref{sec:mp-local} give the size bounds.
Sections~\ref{sec:nn-resolution} and~\ref{sec:mp-weight} control resolution
and weight. Section~\ref{sec:consequences} gives consequences and further
questions. Appendix~\ref{app:dyadic} contains the quantitative geometric
proof. Appendix~\ref{app:mp-alternative} gives an alternative min-plus
weight estimate.

\paragraph{AI methodology.}
The author's 2025 bachelor's thesis at HSE University gave the local
construction with two additional NN anchors per changed input, its
min-plus analogue with real coefficients, and the simultaneous NN
construction with at most $m+3d$ anchors on the Boolean cube.

The present paper adds examples showing that the coefficient $2$ in
the finite-domain size bounds is sharp. It gives a revised resolution
bound for the $m+3d$ NN construction and places all output anchors on a
common dyadic grid (Theorem~\ref{thm:nn-resolution}). For integer
min-plus representations, the paper gives at most $m+d$ forms of weight
at most $W+2n$ on the cube (Theorem~\ref{thm:mp-cube}) and at most
$m+2d$ forms of polynomial weight on finite rational domains
(Theorem~\ref{thm:mp-weight}). Appendix~\ref{app:mp-alternative} gives
an alternative weight bound with at most $m+3d$ forms.

During the subsequent preparation and revision of this manuscript,
ChatGPT (OpenAI) was used to help organize the material, draft and
edit English prose, and proofread the text.
It was also used to review mathematical statements and proofs,
check references, and compare related work. We are responsible for the final
statements, proofs, references, and wording of the manuscript.

\section{Preliminaries}\label{sec:prelim}

Throughout, $n\ge1$. All domains are finite and nonempty, and all functions take values
in $\{0,1\}$. Write $[n]=\{1,\ldots,n\}$, let $\norm{\cdot}$
denote the Euclidean norm, and let $d_H$ be the Hamming distance on
$\cube$. For two functions on the same domain $X$, set
\[
D(f,g)=\{x\in X:f(x)\ne g(x)\},
\qquad \dist_X(f,g)=|D(f,g)|.
\]
We omit the subscript $X$ when $X=\cube$.

We use the NN model of Hajnal, Liu, and Tur\'an~\cite{Hajnal2022},
extending the Boolean-cube definition to arbitrary finite domains.

\begin{definition}[Nearest-neighbor representations]
An NN representation of $f:X\to\{0,1\}$, where $X\subset\R^n$,
is a nonempty finite set of anchors $A\subset\R^n$ together with
labels $\lambda:A\to\{0,1\}$ such that, for every $x\in X$,
\[
\min_{\lambda(a)=f(x)}\norm{x-a}
<\min_{\lambda(b)\ne f(x)}\norm{x-b}.
\]
We use $\min\varnothing=+\infty$. The size is $|A|$, and its
minimum over all representations is denoted by $\NN_X(f)$.
For $X=\cube$ we write $\NN(f)$.
\end{definition}

Ties between nearest anchors with the same label are allowed;
ties between opposite labels at the minimum are forbidden.
Constant functions have complexity one. For functions on the Boolean
cube, restricting the anchors to $\cube$ gives the Hamming
nearest-neighbor complexity $\HNN(f)$. In the NN model the anchors
may lie anywhere in $\R^n$.

The equivalent max-plus formulation was introduced in~\cite{Hansen2015}. We use the
integer-coefficient formulation of~\cite[Definition~5]{DiCicco2024},
extended here to finite rational domains.

\begin{definition}[Integer min-plus representations]
For $X\subset\Q^n$, an integer min-plus representation consists of
two finite families of affine forms with integer coefficients,
$\mathcal L$ and $\mathcal R$, satisfying
\[
f(x)=1 \quad\Longleftrightarrow\quad
\min_{L\in\mathcal L}L(x)\le\min_{R\in\mathcal R}R(x).
\]
We again use $\min\varnothing=+\infty$, and at least one family is nonempty. Its size is
$m=|\mathcal L|+|\mathcal R|$, and its maximum weight $W$
is the largest absolute coefficient, including the constant terms.
The minimum size is denoted by $\MP_X(f)$, or $\MP(f)$ on the cube.
\end{definition}

Thus a form in $\mathcal L$ has label one, a form in $\mathcal R$
has label zero, and a tie between the two minima is resolved in
favor of one. Allowing an empty family makes constants have size one.
For nonconstant functions both families are necessarily nonempty.
Every function on a finite rational domain admits such a representation.
Give each $y\in X$ the form $F_y(x)=\norm{y}^2-2\ip{y}{x}$ with label
$f(y)$. Since $F_y(x)=\norm{x-y}^2-\norm{x}^2$, the unique minimizing
index at $x$ is $y=x$. Clearing all denominators gives integer coefficients.

Following the resolution viewpoint of~\cite{KilicSimaBruck2023},
we measure coordinate precision by bit length and fix the following convention.

\begin{definition}[Resolution]
For a reduced rational number $p/q$, with $q>0$, define
\[
\RES(p/q)=\left\lceil\log_2\bigl(1+\max\{|p|,q\}\bigr)\right\rceil.
\]
The resolution of a rational vector or matrix is the maximum
resolution of its entries. This measures coordinate bit complexity;
the sign requires only one additional bit.
\end{definition}

All logarithms below are to base two. Implicit constants are absolute.
We use the elementary facts that a fixed number of rational arithmetic
operations adds bit lengths, and that integers bounded in absolute
value by $2^{O(L)}$ have $O(L)$ bits. The sum of $n$ rationals
of resolution $R$ has resolution $O(nR+\log(n+1))$. If the summands
have a common denominator of $O(R)$ bits, the corresponding bound is
$O(R+\log(n+1))$.

\section{Local modifications of NN representations}\label{sec:nn-local}

To change the label at a single input, we first put the anchors in
general position.

\begin{lemma}[Generic perturbations]\label{lem:nn-generic}
Let $X\subset\R^n$ be finite and $n\ge2$. Every NN representation
can be perturbed, without changing its size or represented function,
so that the distances from each input to the anchors are pairwise
distinct, and no anchor lies at an input or on a line through two
distinct inputs.
\end{lemma}

\begin{proof}
The strict inequalities between the minimum correct-label distance
and every wrong-label distance persist in an open neighborhood of
the anchor configuration. For each input $x$ and pair of anchor
indices $i\ne j$, the equation
$\norm{x-a_i}^2=\norm{x-a_j}^2$ is the zero set of a nonzero
polynomial in the anchor coordinates and has empty interior.
An anchor lying at an input or on a specified line is also a
condition with empty interior when $n\ge2$. The same applies to
coincident anchors. These forbidden sets are closed with empty
interior, so their finite union cannot cover the open neighborhood
of valid representations. A suitable perturbation therefore exists.
\end{proof}

\begin{theorem}[NN stability]\label{thm:nn-local}
Let $X\subset\R^n$ be finite, $n\ge2$, and let $f,g:X\to\{0,1\}$.
Then
\[
|\NN_X(f)-\NN_X(g)|\le2\dist_X(f,g).
\]
In particular, this bound holds on the Boolean cube for every $n\ge1$.
\end{theorem}

\begin{proof}
It suffices to change the value at one input $y$, using at most two
additional anchors. Apply Lemma~\ref{lem:nn-generic}, and let $a$
be the unique nearest anchor to $y$. Choose $r>0$ so small that
replacing $a$ by any point in $B_r(a)$ preserves, at every input,
all comparisons between its distance to $a$ and its distances to
the other anchors. If $a$ is the only anchor, take any $r>0$.

The vector $y-a$ is nonzero, and no $x-a$, $x\ne y$, is parallel
to it. Therefore there is $v\ne0$ such that
\[
\ip{v}{y-a}=0,
\qquad \ip{v}{x-a}\ne0 \quad(x\in X\setminus\{y\}).
\]
Indeed, the forbidden orthogonality conditions define finitely many
proper subspaces of $(y-a)^\perp$, a space of dimension at least one.
Choose $\varepsilon>0$ sufficiently small that
\[
\varepsilon\norm{v}<r,
\qquad
\varepsilon\norm{v}^2
<2|\ip{v}{x-a}| \quad(x\ne y).
\]
Relabel $a$ by $1-f(y)$ and add $a+\varepsilon v$ and
$a-\varepsilon v$, both with its original label $f(y)$.

At $y$, the squared distance to each new anchor is
$\norm{y-a}^2+\varepsilon^2\norm{v}^2$, so $a$ remains uniquely
nearest and its new label gives the required value. At any $x\ne y$,
\[
\min_{\sigma\in\{-1,1\}}\norm{x-a-\sigma\varepsilon v}^2
=\norm{x-a}^2+\varepsilon^2\norm{v}^2
 -2\varepsilon|\ip{v}{x-a}|
<\norm{x-a}^2.
\]
If $a$ was nearest to $x$, a new anchor with its original label is
now nearest. If another anchor was nearest, the choice of $r$
preserves that fact. Thus only the value at $y$ changes.

Iterating over $D(f,g)$ proves
$\NN_X(g)\le\NN_X(f)+2\dist_X(f,g)$; exchanging $f$ and $g$
gives the absolute-value bound. For the one-dimensional Boolean
cube there are only two inputs, and the bound follows directly
from $1\le\NN(f)\le2$.
\end{proof}

\begin{remark}
The general-domain argument uses the ambient assumption $n\ge2$;
it does not assert the same construction for arbitrary subsets of
$\R$. A collinear domain embedded in $\R^2$ is covered, because the
anchors are free to leave the line.
\end{remark}

\begin{proposition}[Sharpness on finite domains]\label{prop:nn-sharp}
The constant $2$ in Theorem~\ref{thm:nn-local} is optimal.
\end{proposition}

\begin{proof}
Take $X=\{(-1,0),(0,0),(1,0)\}$, let $f$ be constant zero,
and let $g$ equal one only at $(0,0)$. Then $\NN_X(f)=1$.
Two oppositely labeled anchors classify by a strict affine
inequality, which cannot separate the midpoint from both endpoints.
Hence $\NN_X(g)\ge3$, while the three input points themselves
give a representation of size three. Here $\dist_X(f,g)=1$.
\end{proof}

\begin{corollary}\label{cor:nn-sparse}
For $X\subset\R^n$, $n\ge2$, and $k=\min\{|f^{-1}(0)|,|f^{-1}(1)|\}$,
we have $\NN_X(f)\le2k+1$. The same holds on $\cube$ for $n\ge1$.
\end{corollary}

\begin{proof}
Compare $f$ with the closer constant function, which has one anchor.
\end{proof}

On the Boolean cube, the stronger bound $\NN(f)\le k+1$ with constant
coordinate resolution follows from~\cite[Theorem~32]{DiCicco2024}.
For $k>0$, the indicator of the $k$ inputs with the less frequent label
has a DNF with one term per input. If that label is zero, swapping the
anchor labels gives a representation of $f$.

\section{Local modifications of min-plus representations}\label{sec:mp-local}

\subsection{One additional form per changed Boolean input}

The Boolean cube admits an explicit affine separator for each of
its vertices.

\begin{theorem}[Simultaneous modifications on the cube]\label{thm:mp-cube}
Let $f,g:\cube\to\{0,1\}$ and $d=\dist(f,g)$. If $f$ has an
integer min-plus representation $\mathcal F$ with $m$ forms and
maximum weight $W$, then $g$ has such a representation $\mathcal G$
with
\[
|\mathcal G|\le m+d,
\qquad W(\mathcal G)\le W+2n.
\]
The original forms and their labels are retained. Consequently,
\[
|\MP(f)-\MP(g)|\le d.
\]
\end{theorem}

\begin{proof}
For each $y\in D(f,g)$, choose an original form $F_y$ of label
$f(y)$ attaining the global minimum at $y$. Such a form exists
under our tie convention. Define
\[
h_y(x)=2d_H(x,y)-1
=-1+2\sum_{i=1}^n\bigl(y_i+(1-2y_i)x_i\bigr),
\qquad G_y=F_y+h_y,
\]
and give $G_y$ the label $g(y)$. On the cube, $h_y(y)=-1$ and
$h_y(x)\ge1$ for $x\ne y$. Hence $G_y$ is strictly below every
original form at $y$, but is strictly above $F_y$ at every other
input. The forms can be added simultaneously: for $z\ne y$,
\[
G_z(y)>F_z(y)\ge F_y(y)>G_y(y).
\]
Thus $G_y$ is the unique minimum at each changed input $y$, while
no new form attains the minimum outside $D(f,g)$.

The variable coefficients of $h_y$ have magnitude two, and its
constant term has absolute value at most $2n-1$. This proves the
weight bound, including $n=1$.
Applying the construction to a minimum-size representation and
then interchanging $f$ and $g$ proves the final inequality.
\end{proof}

\begin{corollary}\label{cor:mp-sparse}
If $f:\cube\to\{0,1\}$ has $k$ inputs of its less frequent value,
then $\MP(f)\le k+1$, with a representation of maximum weight at
most $2n$. The coefficient $1$ in the size bound of
Theorem~\ref{thm:mp-cube} is optimal.
\end{corollary}

\begin{proof}
Start with a single zero form labeled by the more frequent value.
For sharpness, changing one value of a constant function gives a
nonconstant function, which requires at least two forms.
\end{proof}

\begin{remark}[The role of the domain]\label{rem:exposed}
The size argument applies more generally when every changed point
$y$ is a vertex of $\operatorname{conv}(X)$ for a finite rational
set $X$. Such a point admits a rational affine form $h_y$ that is
negative at $y$ and positive on $X\setminus\{y\}$; clearing its
denominators gives an integer form with the same signs. Adding
$F_y+h_y$ then works exactly as above.
\end{remark}

\subsection{Size on finite rational domains}

For a finite rational domain, write
\[
Q_X=\min\{Q\in\N:Qx\in\Z^n\text{ for all }x\in X\}.
\]

\begin{lemma}[Removing ties]\label{lem:mp-generic}
An integer min-plus representation with $m$ forms and weight $W$
on $X\subset\Q^n$ has an equivalent representation with the same
number of forms, pairwise distinct values at every input, and
weight at most $(2mQ_X+1)W+2m$.
\end{lemma}

\begin{proof}
Put $T=2mQ_X+1$ and replace every form $F_i$ by $TF_i+\tau_i$.
Choose distinct integer offsets in $[0,m-1]$ for label-one forms
and in $[m,2m-1]$ for label-zero forms. Unequal original values
differ by at least $1/Q_X$, so their order is preserved. Equal
values are separated by the offsets, and cross-label ties are
resolved in favor of label one, as required. The stated weight
bound follows immediately.
\end{proof}

For a single changed input, one can relabel its unique minimizing form,
add two small rational perturbations of that form with its old label,
and clear denominators. For several changes we use the simultaneous
construction in Section~\ref{sec:mp-weight}.

\begin{corollary}[Sharp min-plus size bound]\label{cor:mp-local}
For every finite $X\subset\Q^n$ and $f,g:X\to\{0,1\}$,
\[
|\MP_X(f)-\MP_X(g)|\le2\dist_X(f,g).
\]
The constant $2$ is optimal over finite rational domains.
\end{corollary}

\begin{proof}
Apply Theorem~\ref{thm:mp-weight} to a minimum-size representation
of $f$ to obtain $\MP_X(g)\le\MP_X(f)+2\dist_X(f,g)$.
Interchanging $f$ and $g$ gives the reverse bound.

For sharpness, take $X=\{-1,0,1\}$, let $f$ be constant zero, and
let $g$ equal one only at zero. Then $\MP_X(f)=1$ and the bound just
proved gives $\MP_X(g)\le3$. With two forms, a single affine
inequality would have to separate zero from both endpoints.
Hence $\MP_X(g)=3$.
\end{proof}

\section{Resolution of the NN modifications}\label{sec:nn-resolution}

Sequential applications of Theorem~\ref{thm:nn-local} establish
the size bound but do not by themselves control the accumulated
coordinate precision. We instead isolate all changed inputs
before modifying their labels. Section~\ref{sec:nn-geometry}
explains this geometric construction; Appendix~\ref{app:dyadic}
supplies the quantitative details.

\subsection{Resolution and common denominators}

For a rational anchor $a\in\Q^n$, let
\[
q(a)=\min\{q\in\N:qa\in\Z^n\}.
\]
For an anchor matrix $A$ with rows $a_1,\ldots,a_m$, define
\begin{equation}\label{eq:denominator-bits}
\Delta(A)=\max_{1\le i\le m}\left\lceil\log_2\bigl(1+q(a_i)\bigr)\right\rceil.
\end{equation}
A different denominator is allowed for each row.
If $R=\RES(A)$, then $1\le\Delta(A)\le nR$.
In particular, $\Delta(A)=1$ for integer anchors and
$\Delta(A)\le R$ for dyadic anchors. We retain the usual coordinate
resolution $R$ and record $\Delta(A)$ separately.

\Needspace{12\baselineskip}
\begin{theorem}[Resolution under local modifications]\label{thm:nn-resolution}
Suppose $f:\cube\to\{0,1\}$ has a rational NN representation with
anchor matrix $A\in\Q^{m\times n}$. Put $R=\RES(A)$ and
$\Delta=\Delta(A)$. For every $g:\cube\to\{0,1\}$ with
$d=\dist(f,g)$, there is an NN representation with anchor matrix
$B\in\Q^{m'\times n}$ such that
\[
m'\le m+3d,
\qquad
\RES(B)=O\bigl(R+\Delta+\log(n+1)\bigr).
\]
Moreover, all entries of $B$ may be chosen with one common
denominator $2^K$, where $K=O(R+\Delta+\log(n+1))$.
\end{theorem}

\begin{proof}
Assume $n\ge2$. For a Boolean input $x$ and an anchor $a_i$,
the squared distance $\norm{x-a_i}^2$ has denominator dividing
$q(a_i)^2$. Hence every nonzero difference of two squared distances
has absolute value at least
\begin{equation}\label{eq:denominator-gap}
\gamma_0=2^{-4\Delta}.
\end{equation}
In particular, this bound applies to the gap between a nearest
correctly labeled anchor and any incorrectly labeled anchor.

Let $M=n(2^R+2)$, which bounds all input--anchor distances.
For any displacement $e$,
\begin{equation}\label{eq:distance-perturbation}
\left|\norm{x-a-e}^2-\norm{x-a}^2\right|
\le 2M\norm{e}+\norm{e}^2.
\end{equation}
Round the coordinates to a common dyadic grid so that each anchor
moves by at most $\gamma_0/(16(M+1))$. For example, coordinate errors
at most $2^{-K_0}$ suffice for
\[
K_0=\left\lceil\log_2\frac{16\sqrt n(M+1)}{\gamma_0}\right\rceil.
\]
Every squared distance changes
by less than $\gamma_0/4$, so all represented labels are preserved.
Coincident anchors may be removed as in Appendix~\ref{app:nn-general-position}.
A grid exponent $O(R+\Delta+\log(n+1))$ suffices, including the
factor needed to bound Euclidean error from coordinate errors.
Both the grid exponent and the coordinate magnitudes meet the
hypotheses of Lemma~\ref{lem:dyadic} with
$L=O(R+\Delta+\log(n+1))$. That lemma completes the construction.
When $n=1$, use one anchor for a constant function and the two
Boolean inputs for a nonconstant function; these meet the claimed
size and resolution bounds.
\end{proof}

\Needspace{6\baselineskip}
\begin{corollary}\label{cor:nn-rational-resolution}
Under the hypotheses of Theorem~\ref{thm:nn-resolution}, if
$\Delta(A)=O(R)$, the output resolution is $O(R+\log(n+1))$.
Without restrictions on the rational denominators, it is
$O(nR+\log(n+1))$. Both bounds use at most $m+3d$ anchors.
\end{corollary}

\begin{proof}
Substitute $\Delta(A)=O(R)$ or $\Delta(A)\le nR$ into the theorem.
\end{proof}

The first case includes representations whose rows are given as
$a_i=p_i/q_i$, where $p_i\in\Z^n$ and $q_i\in\N$ all have
$O(R)$-bit entries. A separate denominator may be used for each row.
In contrast, bounding individual coordinate denominators alone
does not control their least common multiple by $O(R)$ bits.

\Needspace{12\baselineskip}
\begin{remark}[A margin-sensitive refinement]\label{rem:nn-margin}
For a labeled representation $(A,\lambda)$ of $f$, define
\[
\gamma=\min\left\{1,
\min_{\substack{x\in\cube,\ b\in A\\\lambda(b)\ne f(x)}}
\left(\norm{x-b}^2-
\min_{\lambda(a)=f(x)}\norm{x-a}^2\right)\right\}.
\]
With $\min\varnothing=+\infty$, strict correctness implies
$0<\gamma\le1$. Replacing $\gamma_0$ by $\gamma$ in the rounding
argument gives the potentially sharper output bound
\[
O\bigl(R+\log(n+1)+\log(1/\gamma)\bigr).
\]
This bound also permits a common dyadic output denominator.
\end{remark}

\subsection{Geometry of the simultaneous modification}\label{sec:nn-geometry}

The preceding proof reduces the problem to a common dyadic grid.
The following lemma supplies the simultaneous geometric construction.

\begin{lemma}[Simultaneous construction on a dyadic grid]\label{lem:dyadic}
Suppose an NN representation on $\cube$, $n\ge2$, has $m$
anchors with a common denominator that is a power of two at most
$2^L$, all coordinate magnitudes are at most $2^L$, and
$L\ge\max\{1,\lceil\log_2(n+1)\rceil\}$.
Changing the values at any $d$ inputs admits an NN representation
with at most $m+3d$ anchors and resolution $O(L)$, whose coordinates
share a power-of-two denominator with $O(L)$ bits.
\end{lemma}

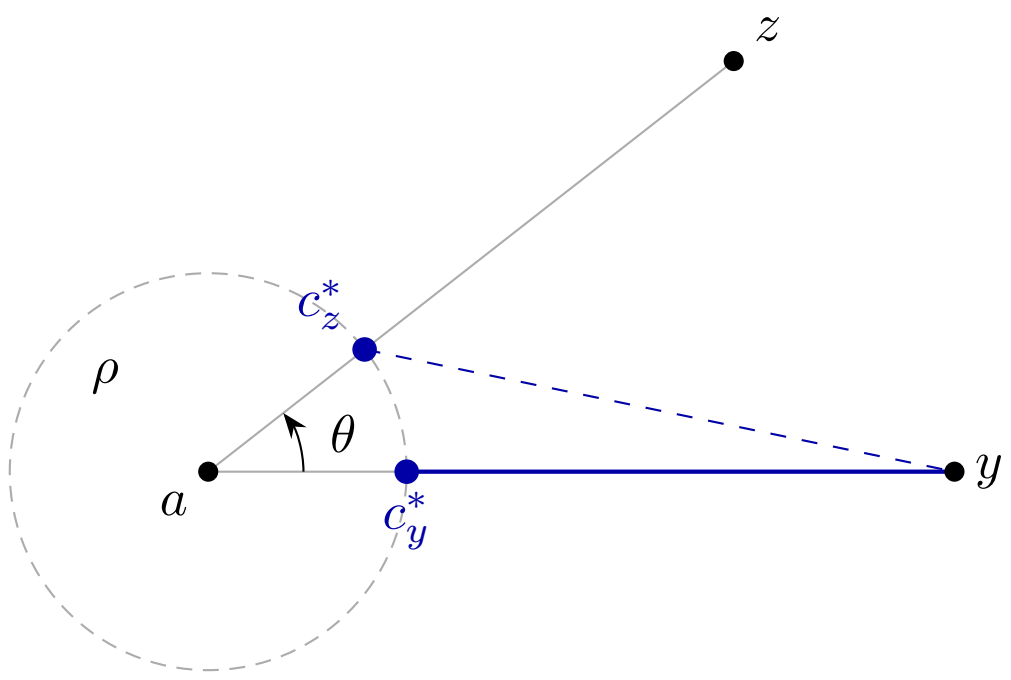
\begin{figure}[htbp]
\centering
\begin{tikzpicture}[scale=1.05,>=Stealth,font=\small]
  \coordinate (a) at (0,0);
  \coordinate (y) at (4.7,0);
  \coordinate (z) at (38:4.2);
  \coordinate (cy) at (1.25,0);
  \coordinate (cz) at (38:1.25);
  \draw[gray!65] (a) -- (y) (a) -- (z);
  \draw[gray!65,densely dashed] (a) circle[radius=1.25];
  \draw[blue!65!black,thick] (cy) -- (y);
  \draw[blue!65!black,dashed] (cz) -- (y);
  \draw[->] (0.6,0) arc[start angle=0,end angle=38,radius=0.6];
  \node at (0.85,0.24) {$\theta$};
  \node at (-0.65,0.6) {$\rho$};
  \fill (a) circle (1.8pt) node[below left] {$a$};
  \fill (y) circle (1.8pt) node[right] {$y$};
  \fill (z) circle (1.8pt) node[above right] {$z$};
  \fill[blue!65!black] (cy) circle (2.2pt) node[below] {$c_y^*$};
  \fill[blue!65!black] (cz) circle (2.2pt) node[above left] {$c_z^*$};
\end{tikzpicture}
\caption{Separating two changed inputs that share a nearest anchor
$a$. Both ideal copies lie at distance $\rho$ from $a$.
The copy on the ray towards $y$ is strictly closer to $y$ than
the other copy. The angle bound makes the gap large enough
to survive rational rounding.}
\label{fig:nn-auxiliary}
\end{figure}

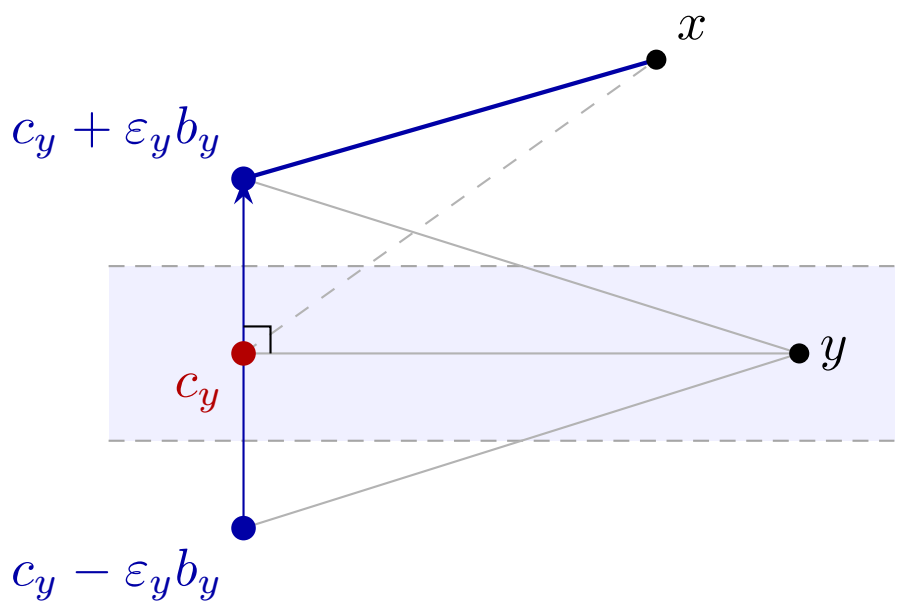
\begin{figure}[htbp]
\centering
\begin{tikzpicture}[scale=1.05,>=Stealth,font=\small]
  \fill[blue!6] (-0.85,-0.55) rectangle (4.1,0.55);
  \draw[gray!70,densely dashed] (-0.85,0.55) -- (4.1,0.55);
  \draw[gray!70,densely dashed] (-0.85,-0.55) -- (4.1,-0.55);
  \coordinate (c) at (0,0);
  \coordinate (plus) at (0,1.1);
  \coordinate (minus) at (0,-1.1);
  \coordinate (y) at (3.5,0);
  \coordinate (x) at (2.6,1.85);
  \draw[gray!60] (c) -- (y) (plus) -- (y) (minus) -- (y);
  \draw[gray!60,dashed] (c) -- (x);
  \draw[blue!65!black,thick] (plus) -- (x);
  \draw[->,blue!65!black] (c) -- (plus);
  \draw[blue!65!black] (c) -- (minus);
  \draw (0,0.17) -- (0.17,0.17) -- (0.17,0);
  \fill[blue!65!black] (plus) circle (2.2pt)
    node[above left] {$c_y+\varepsilon_y b_y$};
  \fill[blue!65!black] (minus) circle (2.2pt)
    node[below left] {$c_y-\varepsilon_y b_y$};
  \fill[red!70!black] (c) circle (2.2pt) node[below left] {$c_y$};
  \fill (y) circle (1.8pt) node[right] {$y$};
  \fill (x) circle (1.8pt) node[above right] {$x$};
\end{tikzpicture}
\caption{Two old-label guards around a relabeled center.
Dashed lines mark the bisectors. For sufficiently small
$\varepsilon_y$, the shaded strip contains $y$ and no other Boolean
input, so each other input is closer to a guard than to the center.
Schematic planar view.}
\label{fig:nn-guards}
\end{figure}

We explain the two main steps in Figures~\ref{fig:nn-auxiliary}
and~\ref{fig:nn-guards}. Appendix~\ref{app:dyadic} gives the full
proof, including the choices of precision. First, a small dyadic
perturbation preserves the represented labels and ensures that no
anchor lies on a line through two Boolean inputs.

\paragraph{One auxiliary anchor per changed input.}
Choose a nearest anchor $a_y$ for each changed input $y$, and move
a copy of $a_y$ a common distance $\rho$ towards $y$:
\[
c_y^*=a_y+\rho\frac{y-a_y}{\norm{y-a_y}}.
\]
The copies initially retain their parents' labels. A sufficiently
small $\rho$ preserves the function and makes $c_y^*$ uniquely
nearest to $y$. To see the separation in the potentially delicate
case, suppose $a_z$ is also nearest to $y$, and let $\theta$ be
the angle between $y-a_z$ and $z-a_z$. Then
\[
\norm{y-c_z^*}^2-\norm{y-c_y^*}^2
=2\rho\norm{y-a_z}(1-\cos\theta)>0.
\]
Noncollinearity gives strict positivity. The Gram-determinant
estimate in Appendix~\ref{app:nn-separation} quantifies this gap,
allowing the ideal copies to be rounded to rational anchors $c_y$
with $O(L)$ bits per coordinate while preserving their separation.

\FloatBarrier

\paragraph{Two guards preserve the other labels.}
For each $y$, choose a rational direction $b_y$ such that
\[
\ip{b_y}{y-c_y}=0,
\qquad \ip{b_y}{x-c_y}\ne0\quad(x\in\cube,\ x\ne y).
\]
Relabel $c_y$ by $g(y)$ and add $c_y\pm\varepsilon_y b_y$
with its old label $f(y)$. Orthogonality makes both guards
farther from $y$ than $c_y$. For $x\ne y$, the squared-distance
identity in Theorem~\ref{thm:nn-local} shows that one guard is
strictly closer than $c_y$ whenever
\[
\varepsilon_y\norm{b_y}^2
<2|\ip{b_y}{x-c_y}|.
\]
Thus a sufficiently small displacement prevents the relabeled
center from winning at any other input. It also keeps guards
from more distant centers from interfering. The quantitative
bounds are proved in Appendix~\ref{app:dyadic}.

The construction uses at most $m+d+2d=m+3d$ anchors.
Appendix~\ref{app:dyadic} bounds its parameters by $O(L)$ bits
using a fixed number of approximation stages.

\Needspace{10\baselineskip}
\section{Weight control on finite rational domains}\label{sec:mp-weight}

We modify the representation simultaneously at all designated inputs.
An integer-valued projection orders the inputs in each cell of the
original representation.
Fix $X\subset\Q^n$ and write
\[
N=|X|,\qquad Q=Q_X,\qquad
B=\max\left\{1,\left\lceil\max_{x\in X}\lVert x\rVert_\infty\right\rceil\right\}.
\]

\begin{theorem}[Simultaneous modifications with bounded weight]\label{thm:mp-weight}
Suppose $f:X\to\{0,1\}$ has an integer min-plus representation with
$m$ forms and maximum weight $W$. Let $g:X\to\{0,1\}$ and
$d=\dist_X(f,g)$. Put
\[
 H=1+d(N-d),\qquad T=nQBH,\qquad P=2d(4T+1).
\]
Then $g$ has an integer min-plus representation $\mathcal G$ with at most
$m+2d$ forms and
\begin{equation}\label{eq:mp-weight}
 W(\mathcal G)\le (QP+1)\bigl((2mQ+1)W+2m\bigr)+P .
\end{equation}
\end{theorem}

\begin{proof}
Apply Lemma~\ref{lem:mp-generic} to obtain integer forms
$\widehat F_1,\ldots,\widehat F_m$ of weight at most
$W_0=(2mQ+1)W+2m$ with pairwise distinct values at each input.
Let $S_j$ consist of the inputs where $\widehat F_j$ is the minimum.
The nonempty $S_j$ partition $X$, and $f$ is constant on each $S_j$.

\paragraph{Ordering the inputs.}
Put $D=D(f,g)$. There is a vector $u\in\{1,\ldots,H\}^n$ such that
$\ip{u}{y-z}\ne0$ for every $y\in D$ and $z\in X\setminus D$.
Indeed, each pair excludes at most $H^{n-1}$ grid vectors, since
one coordinate is determined by the others. There are $d(N-d)=H-1$
pairs, so fewer than $H^n$ vectors are excluded.
The scores $t(x)=Q\ip{u}{x}$ are integers satisfying $|t(x)|\le T$.

Order each nonempty $S_j$ by increasing $t$, breaking ties arbitrarily.
Tied inputs in $S_j$ are either both changed or both unchanged. Since
their original labels agree, their target labels also agree.
Split the ordered inputs into runs on which $g$ is constant.
Write $r_j$ for the number of transitions and $d_j=|D\cap S_j|$.
Each transition has exactly one changed endpoint. Each changed input
has at most two neighbors in the order, so $r_j\le2d_j$.

\paragraph{One form per run.}
At a transition from $z$ to the next input $z'$, we have
$t(z)<t(z')$. Insert the odd integer $2t(z)+1$ between $2t(z)$ and
$2t(z')$, and denote these cuts in increasing order by
$a_{j,1},\ldots,a_{j,r_j}$. Define integer affine forms
\[
 L_{j,0}(x)=0,\qquad
 L_{j,h}(x)=\sum_{s=1}^{h}\bigl(a_{j,s}-2t(x)\bigr)
 \quad (1\le h\le r_j).
\]
Number the runs from $0$ to $r_j$ and give $L_{j,h}$ the label of run
$h$. For $x$ in run $h$, the successive differences
$L_{j,s}(x)-L_{j,s-1}(x)=a_{j,s}-2t(x)$ are negative for $s\le h$
and positive for $s>h$. Thus
$L_{j,h}$ is the unique minimum in this family. This also covers a
cell with a single run, where $L_{j,0}$ is the only form.

On all of $X$, each summand has absolute value at most $4T+1$.
Its constant term has magnitude at most $2T+1$, and each variable
coefficient has magnitude at most $2QH\le2T$. Since $r_j\le2d$,
both the weight of $L_{j,h}$ and $|L_{j,h}(x)|$ for every $x\in X$
are at most $P$.

\paragraph{Combining the cells.}
Put $C=QP+1$. For each nonempty cell use
\[
 G_{j,h}=C\widehat F_j+L_{j,h},\qquad 0\le h\le r_j,
\]
with the labels just assigned. For $x\in S_j$ and $k\ne j$,
$\widehat F_k(x)-\widehat F_j(x)\ge1/Q$, since the forms have integer
coefficients and $Qx\in\Z^n$. Since $L_{j,0}=0$ and
$L_{k,h}(x)\ge-P$, every form from a different nonempty cell satisfies
\[
 G_{k,h}(x)-G_{j,0}(x)\ge C/Q-P=1/Q>0 .
\]
The minimizing form therefore has base index $j$, and the run containing
$x$ gives it the label $g(x)$. The number of forms is at most
$\sum_{j:S_j\ne\varnothing}(1+r_j)\le m+2d$.
Each has weight at most $CW_0+P$, as claimed.
\end{proof}

\begin{corollary}\label{cor:mp-bits}
The representation in Theorem~\ref{thm:mp-weight} satisfies
\[
\begin{aligned}
\log\bigl(W(\mathcal G)+1\bigr)=O\bigl(&\log(W+1)+\log(m+1)+\log(d+1)\\
&+\log(N+1)+\log(n+1)+\log(Q+1)+\log(B+1)\bigr).
\end{aligned}
\]
\end{corollary}

\begin{proof}
Take logarithms in~\eqref{eq:mp-weight} and use that its
right-hand side is a polynomial of fixed degree in the listed
nonnegative parameters.
\end{proof}

Theorem~\ref{thm:mp-weight} bounds the weight polynomially in the
numerical values of the parameters. Corollary~\ref{cor:mp-bits}
bounds the bit length of the weight linearly in their bit lengths.
Since $N=2^n$ on the cube, we use Theorem~\ref{thm:mp-cube} for
the polynomial-weight class below.
For fixed $d$, the bound in~\eqref{eq:mp-weight} is affine in $N$.
Its dependence on $d$ includes the term $d^2(N-d)$.
Proposition~\ref{prop:mp-weight-alternative} in Appendix~\ref{app:mp-alternative}
uses at most $m+3d$ forms and gives a weight bound affine in $d$ for a fixed
domain and input representation.

\section{Consequences and further questions}\label{sec:consequences}

We use the class notation of~\cite[Definitions~8--9]{DiCicco2024}.
For families of Boolean functions $f=(f_n)_{n\ge1}$, let
$\mathsf{NN}$ denote the class admitting polynomially many
real anchors. Define $\mathsf{mpPTF}(\infty)$ similarly
using integer forms, and let $\mathsf{mpPTF}(\poly(n))$
additionally require polynomially bounded maximum weight.

\Needspace{10\baselineskip}
\begin{corollary}[Stability under polynomially many changes]\label{cor:classes}
Let $f_n,g_n:\{0,1\}^n\to\{0,1\}$ satisfy
$\dist(f_n,g_n)\le\poly(n)$. Then, for each of the classes
\[
\mathsf{NN},\qquad
\mathsf{mpPTF}(\infty),\qquad
\mathsf{mpPTF}(\poly(n)),
\]
membership of $f$ is equivalent to membership of $g$.
The same statement holds for NN representations with both
polynomial size and polynomial coordinate resolution.
\end{corollary}

\begin{proof}
Use Theorem~\ref{thm:nn-local} for NN size and
Theorem~\ref{thm:mp-cube} for min-plus size and weight.
Corollary~\ref{cor:nn-rational-resolution} preserves polynomial
resolution. Each argument applies in both directions.
\end{proof}

The additive bounds also transfer lower bounds. For example,
if $\NN(f_n)\ge L(n)$ and $g_n$ differs from $f_n$ on at
most $d(n)$ inputs, then $\NN(g_n)\ge L(n)-2d(n)$.

Theorem~\ref{thm:nn-resolution} also preserves logarithmic
resolution whenever both $R$ and $\Delta(A)$ are $O(\log n)$.
Equivalently, this applies when every initial anchor can be stored
using polynomially bounded integer numerators and one polynomially
bounded denominator. The denominators may differ across anchors.
The output can use a single polynomially bounded dyadic denominator.
Remark~\ref{rem:nn-margin} gives another sufficient condition in
terms of the classification margin.

Several quantitative questions remain. The factor two for NN
size is sharp on arbitrary finite domains, but the example
in Proposition~\ref{prop:nn-sharp} does not establish sharpness
on the full Boolean cube. It is natural to ask whether the
coefficient can be improved there, as it can for min-plus
representations. A second question is whether the NN size
bound $m+2d$ can be combined with a uniform precision bound
as strong as that of the simultaneous $m+3d$ construction.
For an NN anchor $a$, the part of its squared distance to $x$ that depends
on $a$ is $\norm{a}^2-2\ip{a}{x}$. The slope fixes the constant term,
whereas Theorem~\ref{thm:mp-weight} places its cuts by choosing the
constant terms separately. For min-plus representations on the cube,
the conversion to NN anchors in~\cite[Lemma~40]{DiCicco2024} uses one
extra dimension.
Another question is whether the dependence on $\Delta(A)$ can be
removed for arbitrary rational input anchors.

The Hamming model does not admit an analogous bound with an
absolute constant per changed input. Indeed, the constant-one
function and $\mathrm{OR}_n(x)=x_1\vee\cdots\vee x_n$ differ only
at $0^n$, whereas
\[
\HNN(1)=1,\qquad \HNN(\mathrm{OR}_n)=n+1.
\]
For $\mathrm{OR}_n$, the only possible zero-labeled anchor is $0^n$,
which must be present. Strict correctness then forces every unit
vector to be an anchor. These $n+1$ anchors suffice. If $x$ has
weight $k\ge1$ and $x_i=1$, the $i$th unit vector is at Hamming
distance $k-1$ from $x$, while $0^n$ is at distance $k$. This is the boundary construction
of~\cite[proof of Proposition~1]{CepekGlisic2025} applied to $\{0^n\}$
with the labels exchanged.

\clearpage
\appendix
\section{Proof of the dyadic construction}\label{app:dyadic}

We prove Lemma~\ref{lem:dyadic}. Throughout this appendix,
quantities of magnitude $2^{O(L)}$ and positive quantities
bounded below by $2^{-O(L)}$ have constants independent of
$m$ and $d$. Only a fixed number of rounds of approximation
will be used. The case $d=0$ needs no modification, so assume $d>0$.

\subsection{General position with controlled precision}\label{app:nn-general-position}

Let the initial common dyadic denominator be $Q$.
At a Boolean input, every squared anchor distance belongs
to $Q^{-2}\Z$. Hence every nonzero squared-distance gap
is at least $Q^{-2}$. Choose an integer $M=2^{O(L)}$,
$M\ge2$, bounding all input--anchor distances and allowing
an additional perturbation of radius one. Perturbing every
anchor by less than $1/(16MQ^2)$ preserves its represented
function, by the squared-distance estimate
\eqref{eq:distance-perturbation}. This assertion concerns labels
and allows ties between anchors with the same label.

Within this neighborhood, put all coordinates on a finer
common dyadic grid and ensure, for each anchor $a$, that
\begin{equation}\label{eq:six-lines}
a_1\notin\{0,1\},\quad a_2\notin\{0,1\},\quad
a_1\ne a_2,\quad a_1+a_2\ne1.
\end{equation}
This requires avoiding only two values for the first
coordinate and four values for the second. Three and five
successive grid choices, respectively, suffice; the grid
can be chosen fine enough that all these choices stay
inside the permitted neighborhood. Its denominator remains
$2^{O(L)}$.

Condition~\eqref{eq:six-lines} excludes every line through
two Boolean vertices. Indeed, the projection of such a line
onto the first two coordinates is either a Boolean vertex
or a line through two vertices of the square. The latter
lines are precisely the two horizontal, two vertical, and
two diagonal lines excluded by~\eqref{eq:six-lines}.
Remove duplicate anchors if they occur: duplicates of the
same label may be merged, while opposite-label duplicates
cannot be nearest at any input and may both be discarded.
The representation still has at most $m$ anchors.

Let $A$ denote the perturbed anchor set, write $q=2^{O(L)}$ for
its common dyadic denominator, and enlarge $M=2^{O(L)}$ if necessary.
No anchor is a Boolean input, so
\begin{equation}\label{eq:grid-distance}
q^{-1}\le\norm{x-a}\le M
\qquad(x\in\cube, a\in A).
\end{equation}

\subsection{A separate nearest anchor for each changed input}\label{app:nn-separation}

For every $y\in D(f,g)$ choose any nearest original anchor
$a_y$. For a common positive dyadic number $\rho$, define
the ideal auxiliary anchor
\[
c_y^*=a_y+\rho\frac{y-a_y}{\norm{y-a_y}},
\]
with label $f(y)$. Choose
\[
0<\rho<\frac{1}{64Mq^2},
\qquad \log(1/\rho)=O(L).
\]
Since $\rho<q^{-1}$, the distance from $y$ to $c_y^*$ is
$\norm{y-a_y}-\rho$. Adding all these anchors preserves
$f$: each copy retains its parent's label and moves a distance
$\rho$, while the squared-distance gap between a nearest
correct-label anchor and any wrong-label anchor is at least $q^{-2}$.

We need a quantitative uniqueness gap at every selected input.
Fix $y$ and compare $c_y^*$ with $c_z^*$ for $z\ne y$.
If $\norm{y-a_z}>\norm{y-a_y}$, their original squared
distances differ by at least $q^{-2}$, so
\[
\norm{y-c_z^*}-\norm{y-c_y^*}
\ge\norm{y-a_z}-\norm{y-a_y}\ge\frac{1}{2Mq^2}.
\]
If the original distances are equal, put
$u=y-a_z$, $v=z-a_z$, and let $\theta$ be their angle.
These vectors are nonparallel by~\eqref{eq:six-lines}.
Their Gram determinant satisfies
\[
G=\norm{u}^2\norm{v}^2-\ip{u}{v}^2
\in q^{-4}\Z_{>0},
\qquad G\ge q^{-4}.
\]
Using~\eqref{eq:grid-distance},
\[
1-\cos\theta\ge\tfrac12\sin^2\theta
\ge\frac{1}{2q^4M^4}.
\]
Consequently,
\[
\begin{aligned}
\norm{y-c_z^*}^2-\norm{y-c_y^*}^2
&=2\rho\norm{u}(1-\cos\theta)\\
&\ge\frac{\rho}{q^5M^4}.
\end{aligned}
\]
Both distances are at most $M+\rho$, so their sum is at most $4M$.
Dividing the squared-distance difference by this sum gives a
distance gap of at least $\rho/(4q^5M^5)$. The gap between
$c_y^*$ and any original anchor at $y$ is at least $\rho$.
These bounds are uniform over all selected pairs.

Now approximate every $c_y^*$ by a dyadic anchor $c_y$ within
\[
\eta=\frac{\rho}{64q^5M^5}.
\]
Let $q_1$ be the least power of two at least
$\max\{q,8\sqrt n/\eta\}$. The original anchors lie on this grid.
For each copy use the grid choices in Appendix~\ref{app:nn-general-position}
to ensure~\eqref{eq:six-lines} in the first two coordinates and round
the others to the nearest grid point. Each coordinate changes by at
most $5/q_1$, so the displacement is less than $\eta$. Since
$\log(1/\eta)=O(L)$ and $\log n\le L$, we have $q_1=2^{O(L)}$.
Rounding can reduce a distance gap by at most $2\eta$, which is
smaller than the uniqueness gap at each $y$. Each rounded copy is
within $\rho+\eta<2\rho$ of its parent, so its squared distance to
any input changes from the parent's by at most $4M\rho+4\rho^2<q^{-2}$.
All original anchors are retained, and therefore no copy with the
wrong label can reach the original minimum. Thus the intermediate representation
$\mathcal A=A\cup\{c_y:y\in D(f,g)\}$ represents $f$, has
at most $m+d$ anchors, and $c_y$ is uniquely nearest to $y$.

\subsection{Small rational directions for the guards}\label{app:nn-guards}

Fix a selected input $y$. Reflect those coordinates in
which $y_i=1$, sending $y$ to the origin and the cube to
itself. Write $c$ for the reflected auxiliary anchor and
$v=-c$. Condition~\eqref{eq:six-lines} is preserved by
these reflections; in particular $v_1,v_2$ are nonzero and
distinct. We seek a rational vector
\[
b=(t,b_2,1,\ldots,1),
\qquad b_2=\frac{-\sum_{i=3}^n v_i-v_1t}{v_2},
\]
with $\ip{b}{v}=0$ and $\ip{b}{x}\ne0$ for every nonzero
Boolean $x$. An empty sum is zero when $n=2$.

If $x_1=x_2=0$, the latter inner product is the positive
integer $\sum_{i=3}^n x_i$. Otherwise its vanishing excludes
at most one value of $t$ for each of the three possibilities
for $(x_1,x_2)$ and each value
$\sum_{i=3}^n x_i\in\{0,\ldots,n-2\}$. The relevant
slopes are $1$, $-v_1/v_2$, and $1-v_1/v_2$; they are
nonzero because $v_1,v_2\ne0$ and $v_1\ne v_2$.
At most $3(n-1)$ values are forbidden. Thus some integer
$t\in\{1,\ldots,3n+1\}$ works.

The coordinates of $v$ have a common denominator $q_1$,
so the numerator and denominator of $b_2$ have $O(L)$ bits.
All other coordinates are integers of $O(\log(n+1))$ bits.
It follows that
\begin{equation}\label{eq:b-bounds}
\RES(b)=O(L),\qquad
\norm{b}^2\le2^{O(L)},\qquad
\min_{x\in\cube\setminus\{0\}}|\ip{b}{x}|\ge2^{-O(L)}.
\end{equation}
For the last assertion, every such inner product has denominator
dividing that of the single nonintegral coordinate $b_2$.

Undo the reflections on the direction vector. We have obtained,
for each $y$, a direction $b_y$ with
\[
\ip{b_y}{y-c_y}=0,
\qquad
\min_{x\ne y}|\ip{b_y}{x-c_y}|\ge2^{-O(L)},
\]
and the same bounds on resolution and squared norm as
in~\eqref{eq:b-bounds}.

\subsection{Simultaneous correction and final precision}\label{app:nn-correction}

Every nonzero squared-distance gap between anchors of
$\mathcal A$ at a Boolean input is at least $q_1^{-2}$.
Choose a power of two $M_1=2^{O(L)}$, $M_1\ge2$, bounding
all these distances. Every nonzero ordinary distance gap is
at least $1/(2M_1q_1^2)$. Thus the dyadic radius
$r_0=1/(16M_1q_1^2)$ is smaller than one quarter of every
such gap; it also works if there are no nonzero gaps.

For each $y$, choose a positive dyadic $\varepsilon_y$ with
$\log(1/\varepsilon_y)=O(L)$ so small that
\[
\varepsilon_y\norm{b_y}<r_0,
\qquad
\varepsilon_y\norm{b_y}^2
<2\min_{x\ne y}|\ip{b_y}{x-c_y}|.
\]
Relabel $c_y$ by $g(y)$, and add
$c_y\pm\varepsilon_yb_y$, each labeled by $f(y)$.

At $y$, the center $c_y$ is closer than both of its guards,
by orthogonality. Since it was uniquely nearest in
$\mathcal A$, the choice of $r_0$ keeps every other center
and its guards farther away. At an unchanged input $x$,
every nearest center in $\mathcal A$ had label $f(x)$.
An unmodified nearest center keeps this label. A relabeled
nearest center $c_y$, where necessarily $y\ne x$, has a
guard strictly closer to $x$ and carrying its original
label $f(x)$, by the squared-distance identity in the
proof of Theorem~\ref{thm:nn-local}. Descendants of centers
that were not nearest remain farther than an original
minimum-distance center. This proves correctness even
when several original nearest centers were tied.

There are at most $m+d+2d=m+3d$ anchors, each with coordinate
resolution $O(L)$. We also control a common denominator within
each anchor. The center $c_y$ and the number $\varepsilon_y$ are
dyadic, while the direction $b_y$ has integer coordinates except
possibly its second coordinate. If $\varepsilon_y$ has denominator
$2^{k_y}$ and that coordinate of $b_y$ has denominator $s_y$, a
denominator for each guard is $q_1 2^{k_y}s_y$. Consequently every
resulting anchor has a common denominator of $O(L)$ bits.

At a Boolean input, squared distances to anchors with row denominators
$s$ and $t$ differ by a multiple of $1/(s^2t^2)$. If $D_*$ is the
largest row denominator, every positive gap is at least
$\gamma_*=D_*^{-4}$. Let $M_2=2^{O(L)}$ bound all input--anchor
distances. Round the coordinates to a grid of step $2^{-K}$ with
$\sqrt n\,2^{-K}\le\gamma_*/(16(M_2+1))$.
By~\eqref{eq:distance-perturbation}, each squared distance changes by
less than $\gamma_*/4$, preserving the labels. Since
$\log D_*,\log M_2,\log n=O(L)$, we can take $K=O(L)$.
The rounded coordinates have resolution $O(L)$. Coincident anchors
may be removed as in Appendix~\ref{app:nn-general-position}.

\clearpage
\section{An alternative min-plus weight bound}\label{app:mp-alternative}

We use the parameters $N,Q,B$ of Section~\ref{sec:mp-weight}.
The following construction gives a bound with a different dependence
on the number of changed inputs.

\begin{proposition}[Weight bound with three additional forms per change]\label{prop:mp-weight-alternative}
Suppose $f:X\to\{0,1\}$ has an integer min-plus representation
with $m$ forms and maximum weight $W$. For every $g:X\to\{0,1\}$
with $d=\dist_X(f,g)$ there is a representation $\mathcal G$ with
at most $m+3d$ forms. Put
\[
R_*=(d+1)(3nQ^2B^2+2),\qquad M_*=1+4nNQB.
\]
Then the construction satisfies
\begin{equation}\label{eq:mp-weight-alternative}
\begin{aligned}
W(\mathcal G)\le{}&
(QM_*+1)(2QR_*+1)\bigl((2mQ+1)W+2m\bigr)\\
&+(QM_*+1)R_*+M_*.
\end{aligned}
\end{equation}
In particular, $W(\mathcal G)\le\poly(W,m,d,N,n,Q,B)$.
\end{proposition}

\begin{proof}
There is nothing to change when $d=0$, so suppose $d\ge1$.
Apply Lemma~\ref{lem:mp-generic}. Denote the resulting forms
by $\widehat F_1,\ldots,\widehat F_m$, their maximum weight
by $W_0\le(2mQ+1)W+2m$, and the index of the unique
minimum at $x$ by $\mu(x)$.

\paragraph{A distinct minimum for each changed input.}
List $D(f,g)=\{y_1,\ldots,y_d\}$, put $z_i=Qy_i\in\Z^n$,
and define
\[
p_i(x)=\norm{z_i}^2-2Q\ip{z_i}{x}-1,
\qquad r_i(x)=(d+1)p_i(x)+i.
\]
Every $p_i(x)$ is an integer on $X$. Consequently $r_i(x)$
is nonzero, and $r_i(x)\ne r_j(x)$ for $i\ne j$, by reduction
modulo $d+1$. Moreover,
\[
r_i(y_i)\le-1,
\qquad
r_j(y_i)-r_i(y_i)=(d+1)\norm{z_j-z_i}^2+j-i>0
\quad(j\ne i).
\]
The absolute values of both the coefficients and the values
on $X$ of $r_i$ are at most $R_*$.
Here distinct $z_i,z_j\in\Z^n$ satisfy $\norm{z_j-z_i}^2\ge1$,
while $j-i\ge1-d$, which justifies the strict inequality above.

Set $C_0=2QR_*+1$. Retain the forms $C_0\widehat F_j$ and
add
\[
H_i=C_0\widehat F_{\mu(y_i)}+r_i
\]
with the label of $\widehat F_{\mu(y_i)}$.
Distinct base forms have value gaps at least $1/Q$ on $X$;
after multiplication by $C_0$, these gaps exceed $2R_*$.
Thus forms associated with different bases retain their
ordering, and all forms associated with the same base have
the same label. The represented function remains $f$.
The residue argument separates the forms within each base
group, and $H_i$ is the unique minimum at $y_i$.
This intermediate representation has $m+d$ forms with
pairwise distinct values and weight
\[
W_1\le C_0W_0+R_*.
\]

\paragraph{Changing the designated minima.}
For every $i$, choose $u_i\in\{1,\ldots,N\}^n$ such that
$\ip{u_i}{x-y_i}\ne0$ for all $x\ne y_i$.
Each forbidden hyperplane contains at most $N^{n-1}$ grid
points, and there are at most $N-1$ such hyperplanes, so a
choice exists. Define the integer affine forms
\[
S_i(x)=Q\ip{u_i}{x-y_i},
\qquad s_i^\pm(x)=1\pm2S_i(x).
\]
At $y_i$, both $s_i^\pm$ equal one. At every other input,
at least one of them is at most $-1$. Their coefficients
and values on $X$ have absolute value at most $M_*$.

Let $C_1=QM_*+1$. Multiply all intermediate forms by $C_1$,
relabel each $C_1H_i$, and add
$C_1H_i+s_i^+$ and $C_1H_i+s_i^-$ with the original label
of $H_i$.
To check simultaneous correctness, fix an input $x$ and
let $F$ be its unique intermediate minimum. Every other
intermediate form $J$ satisfies
\[
C_1J(x)-M_*>C_1F(x),
\]
because $J(x)-F(x)\ge1/Q$. Thus the forms belonging to a
nonminimal intermediate form cannot win against the group
belonging to $F$.
If $x=y_i$, then $F=H_i$ and its relabeled copy beats both
guards, whose increments equal one. If $x\notin D(f,g)$
and $F=H_i$, a guard with the original label is strictly
below $C_1H_i$. If $F$ is not a designated form, it remains
unchanged. These cases prove correctness at every input.

The two stages add $d+2d$ forms, and the final weight is at
most $C_1W_1+M_*$. Substituting the bounds for $W_1$ and
$W_0$ gives~\eqref{eq:mp-weight-alternative}.
\end{proof}

\clearpage
\bibliographystyle{plain}
\bibliography{local_nn_references}
\end{document}